# Spin Hall Effect

**M. I. Dyakonov**
*Université Montpellier II, CNRS, 34095 Montpellier, France*

## 1. Introduction

The Spin Hall Effect originates from the coupling of the charge and spin currents due to spin-orbit interaction. It was predicted in 1971 by Dyakonov and Perel.[1,2] Following the suggestion in Ref. 3, the first experiments in this domain were done by Fleisher's group at Ioffe Institute in Saint Petersburg,[4,5] providing the first observation of what is now called the Inverse Spin Hall Effect. As to the Spin Hall Effect itself, it had to wait for 33 years before it was experimentally observed by two groups in Santa Barbara (US)[6] and in Cambridge (UK).[7] These observations aroused considerable interest and triggered intense research, both experimental and theoretical, with hundreds of publications.

The Spin Hall Effect consists in spin accumulation at the lateral boundaries of a current-carrying conductor, the directions of the spins being opposite at the opposing boundaries, see Fig. 1. For a cylindrical wire the spins wind around the surface. The boundary spin polarization is proportional to the current and changes sign when the direction of the current is reversed.

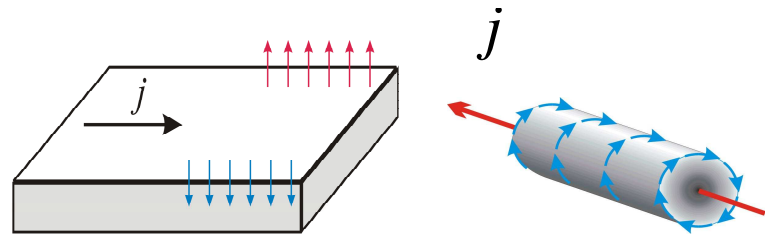

**Figure 1.** The Spin Hall Effect. An electrical current induces spin accumulation at the lateral boundaries of the sample. In a cylindrical wire the spins wind around the surface, like the lines of the magnetic field produced by the current. However the value of the spin polarization is much greater than the (usually negligible) equilibrium spin polarization in this magnetic field.

The term "Spin Hall Effect" was introduced by Hirsch in 1999.[8] It is indeed somewhat similar to the normal Hall effect, where *charges* of opposite signs accumulate at the sample boundaries due to the action of the Lorentz force in magnetic field. However, there are significant differences. First, no magnetic field is needed for spin accumulation. On the contrary, if a magnetic field perpendicular to the spin direction is applied, it will destroy the spin polarization. Second, the value of the spin polarization at the boundaries is limited by spin relaxation, and the polarization exists in relatively wide *spin layers* determined by the spin diffusion length, typically on the order of 1 μm (as opposed to the much smaller Debye screening length where charges accumulate in the normal Hall effect).

Here, we will discuss the phenomenology of spin-charge current coupling, which is true whatever is the microscopic mechanism involved. We will also consider qualitatively the physical origin of various effects resulting from spin asymmetry in electron scattering, in particular the previously unknown phenomenon of *swapping* of spin currents. A more detailed discussion can be found in Ref. 9.

## 2. Phenomenology

We will now discuss, from pure symmetry considerations, what phenomena of spin-charge coupling are in principle possible. Here we restrict ourselves to an isotropic media with inversion symmetry. This does not mean that the results presented below are not valid when inversion symmetry is absent. Rather, it means that we will not take into account additional specific effects, which are due entirely to the lack of inversion symmetry. Our phenomenological approach allows to describe a number of interesting physical effects by introducing only two dimensionless parameters.

### *Spin currents*

The notion of spin current was introduced in Ref. 1. It is described by a tensor $q_{ij}$, where the first index indicates the direction of flow, while the second one says which component of the spin is flowing. Thus, if all electrons with concentration $n$ are completely spin-polarized along $z$ and move with a velocity $v$ in the $x$ direction, the only non-zero component of $q_{ij}$ is $q_{xz} = nv$. (Since $s = 1/2$, it might be more natural to define the spin current density for this case as $(1/2)nv$. It is more convenient to omit 1/2, because this allows avoiding numerous factors 1/2 and 2 in other places. It would be more correct to describe our definition of $q_{ij}$ as the spin *polarization* current density tensor. Below we will use the shorthand "spin current".)

Both the charge current and the spin current change sign under space inversion (because spin is a pseudo-vector). In contrast, they behave differently with respect to time inversion: while the electric current changes sign, the spin current does not (because spin, like velocity, changes sign under time inversion).

*Coupling of spin and charge currents*

The transport phenomena related to coupling of the spin and charge currents can be described phenomenologically in the following simple way.[9,10] We introduce the charge and spin flow densities, $\boldsymbol{q}^{(0)}$ and $q_{ij}^{(0)}$, which would exist in the absence of spin-orbit interaction:

$$q_i^{(0)} = -\mu n E_i - D \frac{\partial n}{\partial x_i}, \qquad (1)$$

$$q_{ij}^{(0)} = -\mu n E_i P_j - D \frac{\partial P_j}{\partial x_i}, \qquad (2)$$

where $\mu$ and $D$ are the mobility and the diffusion coefficient, connected by the Einstein relation, $n$ is the electron concentration, $\boldsymbol{E}$ is the electric field, and $\boldsymbol{P}$ is the vector of spin polarization density (it is convenient to use this quantity, instead of the normal spin density $\boldsymbol{S} = \boldsymbol{P}/2$).

Equation (1) is the standard drift-diffusion expression for the electron flow, while Eq. (2) describes the spin current of polarized electrons, which may exist even in the absence of spin-orbit interaction, simply because spins are carried by the electron flow. We ignore the possible dependence of mobility on spin polarization, which is assumed to be small. If there are other sources for currents, like for example a temperature gradient, the corresponding terms should be included in Eqs. (1) and (2).

Spin-orbit interaction couples the two currents and gives corrections to the values of the primary currents $\boldsymbol{q}^{(0)}$ and $q_{ij}^{(0)}$. For an isotropic material with inversion symmetry, we have:[10]

$$q_i = q_i^{(0)} + \gamma \varepsilon_{ijk} q_{jk}^{(0)}, \qquad (3)$$

$$q_{ij} = q_{ij}^{(0)} - \gamma \varepsilon_{ijk} q_k^{(0)}, \qquad (4)$$

where $q_i$ and $q_{ij}$ are the corrected currents, $\varepsilon_{ijk}$ is the unit antisymmetric tensor (whose non-zero components are $\varepsilon_{xyz} = \varepsilon_{zxy} = \varepsilon_{yxz} = -\varepsilon_{yxz} = -\varepsilon_{zyx} = -\varepsilon_{xzy} = 1$) and $\gamma$ is the small dimensionless parameter proportional to the strength of the spin-orbit interaction. Sums over repeating indices are assumed. The difference in signs in Eqs. (3) and (4) is due to the different properties of charge and spin currents with respect to time inversion.

Thus a $q_{xy}$ spin current induces a charge flow in the $z$ direction ($q_z$), inversely; a charge flow in the $z$ direction induces the spin currents $q_{xy}$ and $q_{yx}$. One can think about this transformation as of a Magnus effect: a spinning tennis ball deviates from its straight path in air in a direction depending on the sense of rotation. So do the electrons, but for another reason, the spin-orbit interaction.

In fact, as it was noticed already in Refs. 1,2, symmetry considerations allow for some additional terms in Eq. (4), to be discussed below.

## 3. Phenomenological equations

Explicit phenomenological expressions for the two currents follow from Eqs. (1)-(4) (the electric current density $j$ is related to $q$ by $j = -eq$, where $e$ is the elementary charge):

$$j/e = \mu n \mathbf{E} + D\nabla n + \beta \mathbf{E} \times \mathbf{P} + \delta \,\mathrm{curl}\, \mathbf{P}, \qquad (5)$$

$$q_{ij} = -\mu E_i P_j - D\frac{\partial P_j}{\partial x_i} + \varepsilon_{ijk}(\beta n E_k + \delta \frac{\partial n}{\partial x_k}). \qquad (6)$$

Here $\beta = \gamma\mu$, $\delta = \gamma D$, so that the coefficients $\beta$ and $\delta$, similar to $\mu$ and $D$, satisfy the Einstein relation. However, since $\gamma$ changes sign under time inversion, $\beta$ and $\delta$ are *non-dissipative* kinetic coefficients, unlike $\mu$ and $D$.

Equations (5) and (6) should be complemented by the continuity equation for the vector of spin polarization:

$$\frac{\partial P_j}{\partial t} + \frac{\partial q_{ij}}{\partial x_i} + \frac{P_j}{\tau_s} = 0, \qquad (7)$$

where $\tau_s$ is the spin relaxation time.

While Eqs. (5)-(7) are written for three dimensions, they are equally applicable to the 2D case, with obvious modifications: the electric field, space gradients, and all currents (but not the spin polarization vector) should have components in the 2D plane only.

## 4. Physical consequences of spin-charge coupling

Equations (5)-(7), which appeared for the first time in Refs. 1, 2, describe the physical consequences of spin-charge current coupling. The effects of spin-orbit interaction are contained in the additional terms with the coefficients $\beta$ and $\delta$.

*Anomalous Hall Effect*

The term $\beta \mathbf{E} \times \mathbf{P}$ in Eq. (5) describes the Anomalous Hall Effect, which is observed in ferromagnets and is known for a very long time (it was first observed by Hall himself in 1881). The measured Hall voltage contains a part, which is proportional to magnetization, but cannot be explained as being due to the magnetic field produced by magnetization (it is much greater than that, especially at elevated temperatures). It took 70 years to understand,[11,12] that the Anomalous Hall Effect is due to spin-orbit interaction.

This effect can also be seen in nonmagnetic conductors, where the spin polarization is created by application of a magnetic field. The spin-related anomalous effect can be separated from the much larger ordinary Hall effect by magnetic resonance of the conduction electrons, which results in a resonant change of the Hall voltage.[13] Non-equilibrium spin polarization produced either by optical means or by spin injection, should also result in an anomalous Hall voltage. Such an experiment was recently done by Miah[14] with GaAs illuminated by circularly polarized light.

*Electric current induced by* **curlP**

The term $\delta$curl**P** in Eq. (5) describes an electrical current induced by an inhomogeneous spin density (now referred to as the Inverse Spin Hall Effect). It can also be regarded as the diffusive counterpart of the Anomalous Hall Effect.

A way to measure this current under the conditions of optical spin orientation was proposed in Ref. 3. The circularly polarized exciting light is absorbed in a thin layer near the surface of the sample. As a consequence, the photo-created electron spin density is inhomogeneous, however curl**P** = 0, since **P** is perpendicular to the surface and it varies in the same direction. By applying a magnetic field parallel to the surface, one can create a parallel component of **P**, thus inducing a non-zero curl**P** and the corresponding surface electric current (or voltage).

This effect was found by Bakun *et al*,[4] providing the first experimental observation of the Inverse Spin Hall Effect, see Fig. 2. In a later publication Tkachuk *et al*[5] observed very clear manifestations of the nuclear magnetic resonance in the surface current induced by curl**P**.

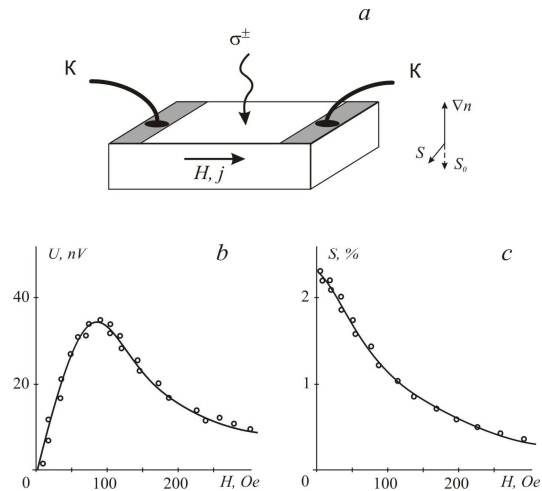

**Figure 2.** First experimental observation of the Inverse Spin Hall Effect. (a) - the experimental setup, (b) - voltage measured between the contacts K as a function of magnetic field, (c) - measured degree of circular polarization of luminescence, equal to the normal component of the average electron spin, as a function of magnetic field. The solid line in (b) is calculated using the results in (c).

*Current-induced spin accumulation, or Spin Hall Effect*

The term $\beta n \varepsilon_{ijk} E_k$ (and its diffusive counterpart $\delta \varepsilon_{ijk} \partial n/\partial x_k$) describes the Spin Hall Effect: an electrical current induces a transverse spin current, resulting in spin accumulation near the sample boundaries.[1] This phenomenon was observed experimentally only in recent years [6,7] and has attracted widespread interest.

Spin accumulation can be seen by solving Eq. (7) in the steady state ($\partial \boldsymbol{P}/\partial t = 0$) and using Eq. (6) for the spin current. Since the spin polarization will be proportional to the electric field, to the first order in $E$ terms $EP$ can be neglected.

We take the electric field along the $x$ axis and look at what happens near the boundary $y = 0$ of a wide sample situated at $y > 0$. The boundary condition corresponds to vanishing of the normal to the boundary component of the spin current, $q_{yj} = 0$.

The solution of the diffusion equation $D d^2 \boldsymbol{P}/dy^2 = \boldsymbol{P}/\tau_s$ with the boundary conditions at $y = 0$, following from Eq. (6), $dP_x/dy = 0$, $dP_y/dy = 0$, $dP_z/dy = \beta n E/D$, gives the result:

$$P_z(y) = P_z(0) \exp(-y/L_s), \quad P_z(0) = -\beta n E L_s/D, \quad P_x = P_y = 0, \qquad (8)$$

where $L_s = (D \tau_s)^{1/2}$ is the spin diffusion length.

Thus the current-induced spin accumulation exists in thin layers (the *spin layers*) near the sample boundaries. The width of the spin layer is given by the spin-diffusion length, $L_s$, which is typically on the order of 1μm. The polarization within the spin layer is proportional to the driving current, and the signs of spin polarization at the opposing boundaries are opposite. These predictions were confirmed by experiments in Refs. 6, 7, and many others.

It should be stressed that all these phenomena are closely related and have their common origin in the coupling between spin and charge currents given by Eqs. (3) and (4). Any mechanism that produces the anomalous Hall effect will also lead to the spin Hall effect and *vice versa*. It is remarkable that there is a single dimensionless parameter, $\gamma$, which governs the resulting physics. The calculation of this parameter should be the objective of a microscopic theory.

*The degree of polarization in the spin layer*

Using Eqs. (9), (11), the *degree* of spin polarization in the spin layer, $\mathrm{P} = P_z(0)/n$, can be rewritten as

$$\mathrm{P} = \gamma \frac{v_d}{v_F} \sqrt{\frac{3\tau_s}{\tau_p}}, \qquad (9)$$

where we have introduced the electron drift velocity $v_d = \mu E$ and used the conventional expression for the diffusion coefficient of degenerate 3D electrons $D = v_F^2 \tau_p /3$, $v_F$ is the Fermi velocity, and $\tau_p$ is the momentum relaxation time. (For

2D electrons, the factor 1/3 should be replaced by 1/2. If the electrons are not degenerate, $v_F$ should be replaced by the thermal velocity.)

In materials with inversion symmetry, like Si, where both the spin-charge coupling and spin relaxation via the Elliott-Yaffet mechanism are due to spin asymmetry in scattering by impurities, the strength of the spin-orbit interaction cancels out in Eq. (9), since $\tau_s \sim \gamma^{-2}$.

Thus the most optimistic estimate for the degree of polarization within the spin layer is $P \sim v_d/v_F$. In semiconductors, this ratio may be, in principle, on the order of 1. In the absence of inversion symmetry, usually the Dyakonov-Perel mechanism makes the spin relaxation time considerably shorter, which is unfavourable for an appreciable spin accumulation. So far, the experimentally observed polarization is on the order of 1%.

*The validity of the approach based on the diffusion equation*

The diffusion equation is valid, when the scale of spatial variation of concentration (in our case, of spin polarization density) is large compared to the mean free path $l = v_F\tau_p$. The variation of $P$ occurs on the spin diffusion length, so the condition $L_s \gg l$ should be satisfied. Since $L_s \sim l(\tau_s/\tau_p)^{1/2}$, this condition can be equivalently re-written as $\tau_s \gg \tau_p$.

Thus, if the spin relaxation time becomes comparable to the momentum relaxation time (which is the case of the so-called "clean limit", when the spin band splitting is greater than $\hbar/\tau_p$), the diffusion equation approach breaks down. The diffusion equation still can be derived for spatial scales much greater than $l$, but it will be of no help for the problem at hand, because neither this equation, nor the boundary conditions for the spin current can any longer be used to study spin accumulation. Surface spin effects will occur on distances less than $l$ from the boundaries and will crucially depend on the properties of the interfaces (e.g. flat or rough interface, etc). To understand what happens near the boundaries, one must address the quantum-mechanical problem of electrons reflecting from the boundary in the presence of electric field and spin-orbit interaction.

## 5. Swapping of spin currents: additional terms in Eq. (4)

Pure symmetry considerations allow for additional terms in one of our basic equations, Eq. (4). Namely, it is possible to complement the rhs of Eq. (4) for the spin current by additional terms of the type: $q_{ji}^{(0)}$ (note the transposition of the indices $i$ and $j$!) and $\delta_{ij}q_{kk}^{(0)}$ (the sum over repeating indices is assumed) with some new coefficients proportional to the spin-orbit interaction.[1,2] This means that spin-orbit interaction will transform the spin currents, for example, turn $q_{xy}^{(0)}$ into $q_{yx}$, i. e. the directions of spin and flow will be interchanged: flow of the $y$ component of spin in the $x$ direction will induce the flow of the $x$ component in the $y$ direction.

This *swapping* of spin currents should lead to new transport effects. For example, suppose that the spins are aligned in the $z$ direction. An electrical current in the *x* direction will be accompanied by the spin current $q_{xz}^{(0)}$, which will induce the spin current $q_{zx}$. Now spins oriented along *x* will flow to the boundaries located in the *z* direction. This will lead to accumulation of *x*-oriented spins at these boundaries resulting in a slight current-induced rotation of the boundary spin polarization around the *y* axis.

The physical origin of spin currents swapping will be discussed in the next section.

## 6. Effects of spin asymmetry in electron scattering

Mott has shown[15,16] that spin-orbit interaction results in an asymmetric scattering of polarized electrons. If a polarized electron beam hits a target, it will deviate in a direction depending on the sign of polarization (similar to a spinning tennis ball in air). This effect is used at high energy facilities to measure the polarization of electron beams (the Mott detectors).

The scattering of electrons by a charged center is schematically depicted in Fig. 3. The most important element for us is the magnetic field ***B*** existing in the electron's moving frame and seen by the electron spin. This field is perpendicular to the plane of the electron trajectory and has opposite signs for electrons moving to the right and to the left of the charged center. The Zeeman energy of the electron spin in this field is, in fact, the spin-orbit interaction.

Simply *looking* at Fig. 3, one can make the following observations:

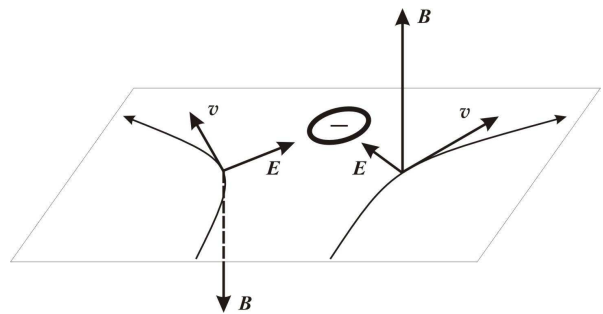

**Figure 3.** Schematics of electron scattering by a negatively charged center. The electron spin sees a magnetic field ***B*** ~ ***v***×***E*** perpendicular to the plane of the electron trajectory. Note that this magnetic field has opposite directions for electrons scattered to the left and to the right.

*Electron spin rotates.* If the electron spin is not exactly perpendicular to the trajectory plane, it will make a precession around ***B*** during the time of collision. The angle of spin rotation during an individual collision depends on the impact parameter and on the orientation of the trajectory plane with respect to spin. This precession is at the origin of the Elliott-Yafet mechanism of spin relaxation.

*The scattering angle depends on spin*. The magnetic field **B** in Fig. 3 is inhomogeneous in space because the electric field **E** is non-uniform and also because the velocity **v** changes along the trajectory. For this reason, there is a spin-dependent force (proportional to the gradient of the Zeeman energy), which acts on the electron. As a consequence, a left-right asymmetry in scattering of electrons with a given spin appears. This is the Mott effect, or *skew scattering*, resulting, among other things, in the Anomalous Hall Effect. If the incoming electrons are *not* polarized, the same spin asymmetry will result in separation of spin-up and spin-down electrons. Spin-ups will mostly go to the right, while spin-downs will go to the left, so that a spin current in the direction perpendicular to the incoming flux will appear (the Spin Hall Effect).

*Spin rotation is correlated with scattering*. As seen from Fig. 3, the spin rotation around the field **B** is correlated with scattering. If the spin on the right trajectory (corresponding to scattering to the right) is rotated clockwise, then the spin on the left trajectory (scattered to the left) is rotated counter-clockwise. The existences of this correlation and its consequences for spin transport[17] have been never discussed previously.

Let us see what happens if the incoming beam ($x$ axis) is polarized along the $y$ axis in the trajectory plane, i.e. characterized by a spin current $q_{xy}$. After scattering, the electrons going to the right will have a spin component along the $x$ axis, while the electrons going to the left will have an $x$ component of the opposite sign! This means that scattering will partly transform the initial spin current $q_{xy}$ to $q_{yx}$. Similarly, $q_{xx}$ will transform to $-q_{yy}$.

Such an analysis shows that during the scattering process the initial spin current $q_{ij}^{(0)}$ generates a new spin current $q_{ij}$ according to the rule:

$$q_{ji}^{(0)} - \delta_{ij} q_{kk}^{(0)} \rightarrow q_{ij}.$$

Thus the correlation between spin rotation and the direction of scattering gives a physical reason for an additional term in Eq. (4), which should be modified as follows:[17]

$$q_{ij} = q_{ij}^{(0)} - \gamma \varepsilon_{ijk} q_k^{(0)} + \kappa (q_{ji}^{(0)} - \delta_{ij} q_{kk}^{(0)}). \quad (10)$$

The last term with the dimensionless coefficient $\kappa$ describes the swapping effect. The effect of skew scattering ($\gamma$) appears only beyond the Born approximation. In contrast, the swapping of spin currents ($\kappa$) is a more robust effect: it exists already in the Born approximation. The swapping coefficient $\kappa$ was calculated in Ref. 17.

In fact, spin-dependent effects in scattering are described by *three* different cross-sections.[17] The first of them describes skew scattering, leading to the Spin Hall Effect, the second one describes the swapping effect, and the third one is responsible for the Elliott-Yafet spin relaxation. These cross-sections were introduced a long time ago by Mott and Massey[16] to analyse the results of a single scattering event in atomic physics.

A possible way to observe the swapping effect[17] is presented in Fig. 4. The primary spin current $q_{yy}^{(0)}$ is produced by spin injection in a semiconductor sample through a ferromagnetic contact. Swapping will result in the appearance of transverse spin currents $q_{xx} = q_{zz} = -\kappa q_{yy}^{(0)}$. Those secondary currents will lead to an excess spin polarization near the lateral boundaries of the semiconductor sample, which could be detected by optical means. At the top face there will be a polarization $P_z < 0$ and at the bottom face $P_z > 0$. (Similarly $P_x < 0$ at the front face and $P_x > 0$ at the back face.) The accumulation of spins polarized perpendicular to the surface distinguishes this manifestation of swapping from the Spin Hall Effect, where the accumulated spins are parallel to the surface.

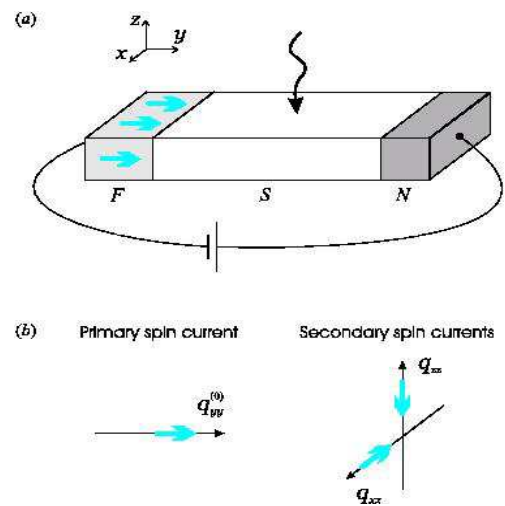

**Figure 4.** Schematics of a proposed experiment revealing the swapping of spin currents. (a) the $q_{yy}^{(0)}$ spin current is electrically injected in a semiconductor (*S*) through a ferromagnetic contact (*F*) with a magnetization along the *y* axis. The wavy arrow symbolizes optical detection of the *z*–component of spin polarization near the surface. (b) the swapping effect transforms the primary spin current $q_{yy}^{(0)}$ into $q_{xx}$ and $q_{zz}$. This should lead to the appearance of excess spin polarization at the lateral boundaries of the sample ($P_z < 0$ at the top face and $P_z > 0$ on the bottom one). This polarization may be detected optically.

## 7. What is it good for, practically?

So far, nobody knows. My personal opinion is that the Spin Hall Effect is one of the many physical phenomena (like for example the Nernst effect, or the Shubnikov-de Haas effect), which do not have any practical applications. Still, it is good that we know and understand them, because such phenomena may help to determine some material parameters and, more importantly, because based on this knowledge we may be able to understand other things, and this eventually might lead to some applications.

However, virtually every paper devoted to this subject start with the standard mantra: "Implementing electrical manipulation of the spin degree of freedom is an essential element for the emerging field of spintronics", or something to this effect. Whether such manipulation (with useful consequences) will ever be possible, or not, still remains to be seen. There are no indications to that so far.

Another issue that is sometimes raised in the current literature, and especially in articles designed for the general public, and which I consider as pure hype, is the "dissipationless spin transport for spin-based electronics". It is proclaimed that dissipationless intrinsic spin currents open the shining perspectives of reducing the power consumption in future spin-based computers, etc.

These statements should be taken with a big grain of salt. First, *all* spin currents by themselves, independently of their microscopic mechanism, are dissipationless. This is simply a consequence of the fact, mentioned in the Section 2, that the spin current, unlike the charge current, does not change sign under time inversion. Related to this, is the property of the "spin Hall mobility", $\beta$, which is a dissipationless kinetic coefficient. Second, the existence of dissipationless spin currents does not mean that one can save energy, because the spin current is induced (with a low efficiency) by the charge current, which *does* involve dissipation. The normal Hall current is dissipationless too. This does not mean that we may reduce energy dissipation in future computers by utilising Hall currents!

## 8. Conclusions

The Spin Hall Effect is a new transport phenomenon, predicted a long time ago but observed only in recent years. It was experimentally studied in three- and two-dimensional semiconductors samples.[6,7,16-21] The Inverse Spin Hall Effect was seen in semiconductors,[4,5,19] as well as in metals.[22,23] Finally, it is important that these effects are observable not only at cryogenic, but also at room temperature.[23,24] However, the number of experimental papers is still about two orders of magnitude less than the number of theoretical ones. At present, it is difficult to predict whether this effect will have any practical applications, as many people believe, or it will belong only to fundamental research as a tool for studying spin interactions in solids.